\documentclass[a4paper]{report}
\usepackage[utf8]{inputenc}
\usepackage[T1]{fontenc}
\usepackage{RJournal}
\usepackage{amsmath,amssymb,array}
\usepackage{booktabs}

\usepackage{longtable}

\makeatletter
 \let\@cite@ofmt\@firstofone
 \def\@biblabel#1{}
 \def\@cite#1#2{{#1\if@tempswa , #2\fi}}
\makeatother
\newlength{\cslhangindent}
\newlength{\csllabelwidth}
 {\begin{list}{}{%
  \setlength{\itemindent}{0pt}
  \setlength{\leftmargin}{0pt}
  \setlength{\parsep}{0pt}
  \ifodd #1
   \setlength{\leftmargin}{\cslhangindent}
   \setlength{\itemindent}{-1\cslhangindent}
  \fi
  \setlength{\itemsep}{#2\baselineskip}}}
 {\end{list}}
\usepackage{calc}

\usepackage{float}
\usepackage{amsmath}
\usepackage{booktabs}
\DeclareUnicodeCharacter{2713}{\checkmark}

\begin{document}

\sectionhead{Contributed research article}
\volume{XX}
\volnumber{YY}
\year{20ZZ}
\month{AAAA}

\begin{article}
\title{A Tidy Data Structure and Visualisations for Multiple Variable Correlations and Other Pairwise Scores}

\author{by Amit Chinwan and Catherine B. Hurley}

\maketitle

\abstract{%
We provide a pipeline for calculating, managing and visualising correlations and other pairwise association scores for numerical and categorical data. We present a uniform interface for calculating a plethora of pairwise scores and propose a tidy data structure for organising the results. We also provide new visualisations which simultaneously show multiple and/or grouped pairwise scores. The visualisations are far richer than a traditional heatmap of correlation scores, as they help identify relationships with categorical variables, numeric variable pairs with non-linear associations or those which exhibit Simpson's paradox. These methods are available in our R package bullseye.
}

\section{Introduction}\label{introduction}

Correlation visualisation, frequently a heatmap matrix, is a common first step in investigating relationships between variables, especially in high-dimensional datasets. Typically, Pearson's correlation for numeric variables is used, but in principle the same visualisations could be used for measures of association which assess non-linear patterns, and for summarising relationships for non-numeric variables. Heatmap visualisations show a single bivariate summary for each pair of variables, but richer visualisations could be designed to show multiple summaries, capturing a variety of bivariate patterns. Such visualisations would also be useful for investigating group-wise correlations and association measures.

In this paper we present a new package \CRANpkg{bullseye} \citep{bullseye} which offers a uniform interface for calculating various scores assessing the relationship between variable pairs, both numerical and categorical. We construct a tidy data structure for managing multiple and/or group-wise bivariate scores. We design a new visualisation method which is a variant of the correlation heatmap matrix for showing multiple pairwise or grouped scores. The variant which compares overall to grouped scores is reminiscent of a bullseye, hence the package name. We further design a new linear display, which is useful for higher-dimensional datasets, when interest lies in displaying just interesting pairs of variables. Both displays are optionally interactive.

More general pairwise variable scores known as scagnostics also fit into the framework offered by \CRANpkg{bullseye}. These numeric variable scores were first introduced by
\citet{tukey1985computer} and further developed by \citet{wilkinson2005graph} and seek to identify patterns such as clumpiness and the presence of outliers, for the purpose of identifying scatterplots worthy of detailed exploration.

The outline of the paper is as follows: In Section 2 we present background, first discussing R packages offering visualisations of correlation summaries, and then move to R packages providing calculation of pairwise variable scores beyond the familiar Pearson, Spearman or Kendall's correlation.
In Section 3 we describe and motivate the choices made in designing a tidy data structure for representing association summaries, and the new displays for their visualisation.
In Section 4 we describe our package \CRANpkg{bullseye}, describing the range of functions for calculating and visualising pairwise associations and give some examples. Section 5 describes how \CRANpkg{bullseye} can be used in co-operation with other R related packages. Finally we conclude with a discussion.

\section{Related R packages}\label{sec:related}

In this section, we provide a brief review of existing R packages used for correlation displays. This establishes the gap that the current work aims to fill. We then provide an overview of methods for calculating pairwise association for numerical and factor variables, specifically with reference to their R package implementations

\subsection{Displaying correlation}\label{displaying-correlation}

There are many available R packages providing correlation displays, the most notable are summarized in Table \ref{tab:corrdisplay-packages}. For each package we list the display type offered, whether categorical variables are supported and whether grouped correlations are on offer. Most of these use a matrix layout of glyphs or filled squares (i.e., a heatmap) and cater for numerical variables only.

\begin{table}[!h]
\centering
\caption{\label{tab:corrdisplay-packages}List of R packages for correlation display, with the display type, whether categorical variables are supported, whether grouped correlations are provided, and whether a dataframe is used to represent correlations.}
\centering
\begin{tabular}[t]{llccc}
\toprule
package & Display type & Categorical & Groups & Dataframe\\
\midrule
ellipse & ellipse matrix &  &  & \\
heplots & ellipse matrix &  & ✓ & \\
corrplot & heatmap/glyph matrix &  &  & \\
corrgram & heatmap/glyph matrix &  &  & \\
ggcorrplot & heatmap/glyph matrix &  &  & \\
corrr & glyph matrix/network &  &  & ✓\\
linkspotter & network & ✓ &  & ✓\\
corrgrapher & network & ✓ &  & \\
corrViz & various &  &  & \\
correlationfunnel & linear & ✓ &  & \\
\bottomrule
\end{tabular}
\end{table}

\citet{murdoch1996} proposed a display for correlation matrices which uses a matrix layout of ellipses where the parameters of the ellipses are scaled to the correlation values. An implementation is provided in package \CRANpkg{ellipse} \citep{ellipse}. \citet{friendly2002corrgrams} expanded on this idea by rendering correlation values as shaded squares, bars, ellipses, or circular pac-man symbols, and these methods are implemented in packages \CRANpkg{corrplot} \citep{corrplot} and \CRANpkg{corrgram} \citep{corrgram}. The \CRANpkg{ggcorrplot} \citep{ggcorrplot} and \CRANpkg{corrr} \citep{corrr} packages also provided heatmap-type displays, but based on \CRANpkg{ggplot2} \citep{ggplot2} rather than base R.
\citet{FriendlySigal} (in their Figure 2) present displays of grouped correlations as overlaid ellipses. This is implemented in the function \texttt{covEllipse} of package \CRANpkg{heplots} \citep{heplots}.

A number of packages support network displays of correlation values. In the package \CRANpkg{corrr} network displays use line thickness to encode correlation magnitude, with a filtering option to discard low-correlation edges.
\CRANpkg{linkspotter} \citep{linkspotter} offers a variety of association measures in addition to correlation, where the measure used depends on whether the variables are both numerical, categorical or mixed. The results are visualised in a network plot, which may be packaged into an interactive shiny application.
The package \CRANpkg{corrgrapher} \citep{corrgrapher} also uses a network plot, and handles mixed type variables by using association measures obtained as transformations of \(p\)-values obtained from Pearson's correlation test in the case of two numeric variables, Kruskal's test for numerical and factor variables, and a chi-squared test for two categorical variables.

The R package \CRANpkg{correlationfunnel} \citet{correlationfunnel} offers a novel linear display which assists in feature selection in a regression setting with many predictor variables. All numeric variables including the response are binned. All (now categorical) variables in the resulting dataset are one-hot encoded and Pearson's correlation calculated with the response categories. The correlations are visualised in a dot-plot display, where predictors are ordered by maximum correlation magnitude, giving a funnel-like shape.

The majority of the listed packages display the result of the standard \texttt{cor} calculation, so the input data structure for the visualisations is a \(p \times p\) symmetric matrix of numerical values, where \(p\) is the number of variables. \CRANpkg{corrr} uses a \(p \times p\) dataframe instead, or alternatively a long format dataframe where each pair of variables form a row, and provides some convenience functions for manipulating these structures. \CRANpkg{linkspotter} similarly uses a long dataframe where pairs of variables form a row, with additional columns for possible multiple correlations. The package \CRANpkg{correlation} \citep{correlationPackage} also uses a long dataframe, and for grouped correlations each row is associated with a variable pair and group level. It provides heatmap and network-style visualisations, though not, as far as we can tell, for grouped correlations.

\subsection{Calculating association measures}\label{calculating-association-measures}

Many alternative measures of association have been devised, beyond the standard Pearson, Spearman or Kendall's correlation. The purpose of these measures is to capture non-linear associations, associations between nominal and ordinal variables, as well as associations between variables of mixed type. Implementations of these measures are scattered across many R packages.

For nominal variable pairs, many classical measures of association are implemented in the R package \CRANpkg{DescTools} \citep{DescTools}. One of these is Pearson's contingency coefficient which uses the \({\chi}^2\) value (scaled to the unit interval) from Pearson's \({\chi}^2\) test for independence. Another is the uncertainty coefficient \citep{theil1970estimation} which measures the proportion of uncertainty in one variable which is explained by the other. A symmetric version of this coefficient is obtained by taking the mean of the uncertainty coefficients treating each variable as an independent variable once.

\CRANpkg{DescTools} also offers implementations of many association measures for ordinal pairs of variables, including Kendall's \(\tau\) and \(\tau_b\) and Goodman Kruskal's gamma, all of which are based on the on the number of concordances and discordances in paired observations (see for example \citet{agresti2010analysis} for a description). \CRANpkg{polycor} \citep{polycor} implements polychoric correlation \citep{olsson1979maximum}, which summarizes the association between two ordinal variables by assuming two normally distributed latent variables. It also provides polyserial correlation, a measure of association between an ordinal, numeric pair of variables.

For numeric pairs of variables, the recently proposed measures distance correlation \citep{szekely2007measuring} and maximal information coefficient (MIC) \citep{reshef2011detecting} are measures of dependence which aim to capture non-linear as well as linear patterns. Distance correlation is calculated as the correlation between two sets of distances, and is available from the R package \CRANpkg{energy} \citep{energy}. The MIC coefficient is calculated as the mutual information between discretised versions of the two numeric variables, where the binning is chosen to maximise the mutual information. \citet{commentSimonTibshirani} simulated pairs of variables with different relationships at varying levels of noise and showed that distance correlation has more statistical power than MIC. In a later study, \citet{kinney} proposed additional mutual information-based measures with improved performance. The MIC and other maximal information based non-parametric exploration statistics have been implemented in the R package \CRANpkg{minerva} \citep{minerva}. Another package \CRANpkg{linkspotter} provides a mutual information measure for pairs of variables which may both be numeric, factors or mixed, which is similar to the \CRANpkg{minerva} MIC measure for numeric pairs.

The alternating conditional expectations (ACE) algorithm \citep{breiman1985estimating} as implemented in \CRANpkg{acepack} \citep{acepack} estimates optimal transformations between response and predictor variables for regression analysis and calculates \(R^2\) using this optimal transformation. For a numeric pair of variables, this provides another way of summarizing their non-linear association. For categorical variables, ACE transformations calculates the optimal scoring of levels (also known as canonical analysis), and the resulting \(R^2\) provides an alternative measure of association.

The package \CRANpkg{scagnostics} \citep{scagnostics} implements the methods in \citet{wilkinson2005graph} evaluating scatterplot ``interestingness'', and provides nine scores called outlying, skewed, clumpy, sparse, striated, convex, skinny, stringy and monotonic.

\section{Bullseye design}\label{sec:design}

In this section we present design choices made in constructing the new package \CRANpkg{bullseye}.

We firstly describe a tidy data structure for managing pairwise scores, which offers far more flexibility than the standard correlation matrix.
While we anticipate most users of the package will be more interested in the visualisations provided, understanding this structure will be of interest to those who wish to use \CRANpkg{bullseye} calculations with visualisations from other packages, or conversely those who wish to use association measures from other packages with \CRANpkg{bullseye} visualisaions.

Then we describe and motivate our new visualisations.

\subsection{Design of tidy data structure}\label{sec:pairwisedesign}

Conventionally, correlation measures are represented as a matrix, following the mathematical definition of a correlation matrix. This matrix representation does not follow tidy data principles as described in \citep{wickhamjss}.
Furthermore, the matrix arrangement does not extend naturally to the situation of multiple or grouped pairwise measures. For these reasons, some existing packages (for example \CRANpkg{corrr}, \CRANpkg{linkspotter} and \CRANpkg{correlation}) provide tidy dataframe representations.

Organising data according to tidy data principles means variables are columns and observations are rows.
For example, a correlation matrix could be converted to a tidy format by having two variables for the row and column label of the matrix, and a third variable for the correlation value.

For our purposes, we use a \texttt{tibble} structure that contains as variables:

\begin{enumerate}
\def\labelenumi{(\roman{enumi})}
\item
  the two dataset variables \texttt{x} and \texttt{y},
\item
  \texttt{score}: the score name identifying the summary measure used for the two variables,
\item
  \texttt{group}: identifying the subset of the data used for calculating the summary and
\item
  \texttt{value}: a numeric value recording the value of the summary measure.
\end{enumerate}

As the pairwise measures we consider are symmetric, each variable pair appears once only. By construction the \texttt{x} and \texttt{y} variables are in alphabetical order, that is, \texttt{x} \textless{} \texttt{y}. Note there is no row where \texttt{x} = \texttt{y}.
An additional column \texttt{pair\_type} is included for information purposes. Here we use \texttt{nn} for two numeric variables, \texttt{ff} for two factors and \texttt{fn} for a factor-numeric pair, or NA if unknown.
The variables \texttt{x}, \texttt{y}, \texttt{score}, \texttt{group} uniquely identify each row, that is,
the pairwise tibble has at most one row for each combination of \texttt{x}, \texttt{y}, \texttt{score} and \texttt{group}. This requirement will be checked prior to using \CRANpkg{bullseye} visualisaions.

\subsection{Design of visualisations}\label{sec:visdesign}

Ideally, we would design a visualisation where each of \texttt{x}, \texttt{y}, \texttt{score}, \texttt{group} and \texttt{value} columns of the pairwise tibble are mapped to a visual channel, that is, some aspect of the plot.
\citet{cf13} identified over 30 visual channels, most of which have not been subjected to detailed comparisons in the literature.
\citet{cleveland84} proposed a ranking of 10 channels from most accurate (position along a common scale) to least accurate (colour saturation), and conducted experiments based on two-value ratio judgements to verify the ranking, though the experiments excluded colour.
\citet{borgo13} reported on perceptual studies where the ``pop-out'' effects of colour are far greater than those of other channels such as size, shape and orientation.
More recently, \citet{coleman22} conducted extensive experiments on six visual channels, including luminance.
For the task of reproducing up to eight values, they found that
the rankings of \citet{cleveland84} did not hold, and that performance is vastly more affected by the number of items to be compared than the channel used. Of the six visual channels explored, luminance was ranked mid table across the various settings.
They concluded that much more work needs to be done to rank visual channels across a range of scenarios.

\subsubsection{Matrix display}\label{sec:visdesignmat}

A natural starting point for the visualisation of the pairwise tibble would be a generalisation of the familiar heatmap or glyph representation of a matrix, such as displays provided by many of the packages listed in Table \ref{tab:corrdisplay-packages}.
Then the \texttt{x} and \texttt{y} variables are mapped to layout positions,
and the other variables \texttt{score}, \texttt{group} and \texttt{value} are mapped to a glyph. \citet{borgo13} and \citet{Ward2008} both give useful surveys of glyph-based design.

In designing the visualisation, note that the pairwise tibble could have:

\begin{enumerate}
\def\labelenumi{\alph{enumi}.}
\item
  one measure and value for each \texttt{x}, \texttt{y} pair,
\item
  multiple measures, groups and values for each \texttt{x},\texttt{y} pair, or
\item
  multiple measures, groups and values for each \texttt{x},\texttt{y} pair, where there is a measure for \texttt{group\ =\ "all"} and others for subsets.
\end{enumerate}

The first case (a) is the simplest, which is addressed by the matrix displays of packages listed in Table \ref{tab:corrdisplay-packages}.
The most popular display is a matrix of filled squares or circles, using a diverging colour scale, where colour hue represents positive or negative values, and
other colour aspects (chroma and luminance) represent the magnitude.
This is our choice of display in \CRANpkg{bullseye} for this setting (see Figure \ref{fig:plot-pairwise-scores}).
Generally, the analyst uses these matrix displays
to easily identify pairs of variables with strong, weak or negative associations, rather than for accurate reading of precise association values, so Cleveland and McGill's hierarchy of visual channels is less of a concern.

For case (b), we decided to split the filled circle into sectors of equal size, one for each value, where each sector is assigned a colour representing a value. See Figure \ref{fig:plot-pairwise-multi-latex} for an example. With this visualisation, the pairs of variables with strong, weak or negative associations, and pairs of variables with large differences across the different values ``pop-out'' strongly.

For the special case of (c) where some measures represent all the data and others represent subsets, we divide the circle into an
inner section whose sectors show ungrouped score values for \texttt{group\ =\ "all"}, and the outer doughnut whose sectors shows the score values for other groups. As before, sectors are assigned a colour representing the value sign and magnitude.
This glyph design represents the hierarchical nature of the values when some summarise the entire dataset and others
summarise the data which has been split into subsets. An example is given in Figure \ref{fig:plot-sc-species-latex}.
An interesting pattern in this context is where the direction of the association is reversed when data is split into groups (that is, Simpson's paradox),
and in our visualisation, this pattern is easily identified by distinct hues for the inner and outer sectors.

\citet{Ward2008} also discusses the role of ordering, or seriation, in constructing more informative glyph displays.
For our matrix displays, there are two possible uses of seriation (i) ordering the (\texttt{x},\texttt{y}) pairs and (ii) ordering the slices within each glyph.
At the moment we do not implement a strategy for (ii).
For (i), we use seriation to order variable pairs to help draw attention to interesting glyphs.
In the case of a single score for each (\texttt{x},\texttt{y}) pair, the usual way of ordering variables employs hierarchical clustering to place high-score (typically correlation) pairs nearby. This has the effect of placing high-score pairs of variables near the diagonal of the matrix layout. We use a variant of this scheme, which additionally aims to give high scoring pairs extra prominence by placing them in the top-left corner, using the \citet{dendserPaper} lazy path length algorithm implemented in package \CRANpkg{DendSer} \citep{DendSer}.
For cases (b) and (c) with multiple scores, interesting pairs of variables will likely be those with one or more high scores, or alternatively those high variation across the scores. In these settings our seriation procedure arranges the variable pairs based on the
the maximum of the absolute values, or alternatively the maximum difference of those values.

In a previous version of this work \citep{chinwan} (see for example Figure 3.4 and 3.6) we experimented with representing values by lines using position along a common scale, the most accurate representation according to \citet{cleveland84}. However, the display did not make good use of available space and in our opinion did not have the same ``pop-out'' effect as our current designs, though this assessment has not been subjected to experimentation.

Note that in the matrix layout with the divided circle glyph the \texttt{group} and \texttt{score} variables are mapped to sector position, making it difficult for the user to see which sector is which. This is a problem common to multivariate glyphs in general, for example stars and Chernoff faces, and as discussed in \citet{Ward2008}, can lead to bias in glyph mappings. In our setting, we have opted to add interactive tool tips to the displays so that the group and or score as well as the precise value associated with each sector is easily identified.

\subsubsection{Linear display}\label{linear-display}

The matrix display described above is suitable when the number of variables is not too large.
As the number of variables increases, the pairwise tibble could be filtered to show interesting (\texttt{x},\texttt{y}) pairs only, or in the case of a regression problem, pairs where a specified response variable is present.
For this setting we use a dotplot display, where (\texttt{x},\texttt{y}) pairs are mapped to rows, \texttt{value} is mapped to position, and the group/score categories are mapped to point colour.
We also offer a heatmap variant where \texttt{value} is mapped to tile fill.
See Figures \ref{fig:plot-sc-multi-latex} and
\ref{fig:plot-group-scores} for examples.

Once again, seriation is used to order variable pairs to help draw attention to high-score pairs. Here variable pairs are ordered by their maximum absolute value, or alternatively the maximum difference.

\section{Bullseye functionality}\label{sec:bullseyefunc}

In this section we describe the main functions of our new package \CRANpkg{bullseye}. First we describe the functions included for calculating individual, multiple and grouped pairwise scores. Then we present visualisation functions and give some examples.

We demonstrate the functions using the \texttt{penguins} dataset from package \CRANpkg{palmerpenguins} \citep{palmerpenguins}, which lists measurements for penguin species from three islands in the Palmer Archipelago. The dataset has five numerical variables, and three factor variables \texttt{species}, \texttt{island} and \texttt{sex}.

\subsection{Calculating pairwise scores}\label{sec:bullseyecalc}

Starting from a correlation matrix of numeric variables in the \texttt{penguins} dataset, constructed below as \texttt{peng\_cor}, the \texttt{pairwise} function constructs the equivalent pairwise tibble:

\begin{verbatim}
peng_cor <- select(penguins, where(is.numeric)) |>
  cor(use= "pairwise.complete.obs")

pairwise(peng_cor, score="pearson", pair_type = "nn")
\end{verbatim}

\begin{verbatim}
#> # A tibble: 10 x 6
#>   x         y         score   group  value pair_type
#>   <chr>     <chr>     <chr>   <chr>  <dbl> <chr>    
#> 1 bill_dep  bill_len  pearson all   -0.235 nn       
#> 2 bill_len  flip_len  pearson all    0.656 nn       
#> 3 bill_dep  flip_len  pearson all   -0.584 nn       
#> 4 body_mass flip_len  pearson all    0.871 nn       
#> 5 bill_len  body_mass pearson all    0.595 nn       
#> 6 bill_dep  body_mass pearson all   -0.472 nn       
#> # i 4 more rows
\end{verbatim}

We also provide utility functions to calculate various pairwise variable scores.
These are listed in Table \ref{tab:pair-fns}.
Each of these functions takes as input a \texttt{data.frame} containing a dataset and \texttt{handle.na} which defaults to \texttt{TRUE} (and uses pairwise complete observations). Each excludes variables which are not of the appropriate type for the score to be calculated. Other arguments may optionally be passed to the underlying calculation function, specified in the ``From'' column of Table \ref{tab:pair-fns}. Each of these \texttt{pair\_*} functions produces an output which is a tibble of class \texttt{pairwise}, with the \texttt{score} and \texttt{pair\_type} columns filled in as appropriate. The function \texttt{pair\_mine} will optionally provide multiple measures for each variable pair, while \texttt{pair\_scagnostics} defaults to providing all nine scagnostics from \CRANpkg{scagnostics}.

\begin{table}[!h]
\centering
\caption{\label{tab:pair-fns}Bullseye functions for calculating pairwise correlations, other association measures and scagnostics. Tick marks in the columns nn, ff, fn indicate whether the function gives a score for numeric, factor and factor-numeric pairs. Ticks in the Ordinal column indicates method assumes ordinal factors.}
\centering
\resizebox{\ifdim\width>\linewidth\linewidth\else\width\fi}{!}{
\begin{tabular}[t]{llccclll}
\toprule
Function & Description & nn & ff & fn & From & Range & Ordinal\\
\midrule
pair\_cor & Pearson/Spearman/Kendall & ✓ &  &  & cor & {}[-1,1] & \\
pair\_dcor & Distance correlation & ✓ &  &  & energy::dcor2d & {}[0,1] & \\
pair\_mine & MIC and other measures & ✓ &  &  & minerva::mine & {}[0,1] & \\
pair\_ace & Ace correlation & ✓ & ✓ & ✓ & acepack::ace & {}[0,1] & \\
pair\_cancor & Canonical correlation & ✓ & ✓ & ✓ & cancor & {}[0,1] & \\
pair\_nmi & MIC & ✓ & ✓ & ✓ & linkspotter::maxNMI & {}[0,1] & \\
pair\_polychor & Polychoric correlation &  & ✓ &  & polycor::polychor & {}[-1,1] & ✓\\
pair\_polyserial & Polyserial correlation &  &  & ✓ & polycor::polyserial & {}[-1,1] & ✓\\
pair\_tauA & Kendall’s tau A &  & ✓ &  & DescTools::KendalTauA & {}[-1,1] & ✓\\
pair\_tauB & Kendall’s tau B &  & ✓ &  & DescTools::KendalTauB & {}[-1,1] & ✓\\
pair\_tauC & Stuart-Kendall Tau-C &  & ✓ &  & DescTools::StuartTauC & {}[-1,1] & ✓\\
pair\_tauW & Kendall’s W &  & ✓ &  & DescTools::KendalW & {}[-1,1] & ✓\\
pair\_gkGamma & Goodman-Kruskall Gamma &  & ✓ &  & DescTools::GoodmanKruskalGamma & {}[-1,1] & ✓\\
pair\_gkTau & Goodman-Kruskall tau &  & ✓ &  & DescTools::GoodmanKruskalTau & {}[0,1] & ✓\\
pair\_uncertainty & Uncertainty coefficient &  & ✓ &  & DescTools::UncertCoef & {}[0,1] & \\
pair\_chi & Pearson's contingency coefficient &  & ✓ &  & DescTools::ContCoef & {}[0,1] & \\
pair\_scagnostics & Scagnostics & ✓ &  &  & scagnostics::scagnostics & {}[0,1] & \\
\bottomrule
\end{tabular}}
\end{table}

The calculation in the previous code snippet can be done in one step with the \texttt{bullseye} function \texttt{pair\_cor}. Notice that this automatically excludes non-numeric variables and handles missing values using pairwise complete observations. In the example below, the optional argument \texttt{method\ =\ "spearman"} is specified so Spearman correlation is calculated, rather than the default Pearson.

\begin{verbatim}
pair_cor(penguins, method="spearman")
\end{verbatim}

\begin{verbatim}
#> # A tibble: 10 x 6
#>   x         y         score    group  value pair_type
#>   <chr>     <chr>     <chr>    <chr>  <dbl> <chr>    
#> 1 bill_dep  bill_len  spearman all   -0.222 nn       
#> 2 bill_len  flip_len  spearman all    0.673 nn       
#> 3 bill_dep  flip_len  spearman all   -0.523 nn       
#> 4 body_mass flip_len  spearman all    0.840 nn       
#> 5 bill_len  body_mass spearman all    0.584 nn       
#> 6 bill_dep  body_mass spearman all   -0.432 nn       
#> # i 4 more rows
\end{verbatim}

Here is \texttt{pair\_mine}, where two methods are specified. The default choice here is just \texttt{method="MIC"}.

\begin{verbatim}
pair_mine(penguins, method=c("MIC", "TIC"))
\end{verbatim}

\begin{verbatim}
#> # A tibble: 20 x 6
#>   x        y         score group value pair_type
#>   <chr>    <chr>     <chr> <chr> <dbl> <chr>    
#> 1 bill_dep bill_len  MIC   all   0.313 nn       
#> 2 bill_dep bill_len  TIC   all   0.236 nn       
#> 3 bill_dep body_mass MIC   all   0.518 nn       
#> 4 bill_dep body_mass TIC   all   0.435 nn       
#> 5 bill_dep flip_len  MIC   all   0.660 nn       
#> 6 bill_dep flip_len  TIC   all   0.510 nn       
#> # i 14 more rows
\end{verbatim}

Additional \texttt{bullseye} functions provide multiple pairwise variable scores and/or grouped scores, and these are listed in Table \ref{tab:pairwise-fns}.

\begin{table}[!h]
\centering
\caption{\label{tab:pairwise-fns}Bullseye functions for combining multiple pairwise scores, and for handling grouping variables. }
\centering
\begin{tabular}[t]{l>{\raggedright\arraybackslash}p{10cm}}
\toprule
Function & Description\\
\midrule
pairwise\_multi & Combines outputs of  pair\_*  functions.\\
pairwise\_by & Combines outputs of  a pair\_*  function for every level of a grouping variable.\\
pairwise\_scores & Combines outputs of  pair\_* functions, where different pair\_* functions may be specified for nn, fn, and ff pairs. A grouping variable may optionally be specified.\\
\bottomrule
\end{tabular}
\end{table}

Dataframes of type \texttt{pairwise} may of course be bound together rowwise. For convenience, we provide another function \texttt{pairwise\_multi} which calculates multiple association measures for every variable pair in a dataset. The example below calculates distance correlation and MIC for penguin variables.

\begin{verbatim}
dcor_nmi <- pairwise_multi(penguins, c("pair_dcor", "pair_nmi"))
\end{verbatim}

For each of the \texttt{pairwise} calculation functions, they can be wrapped using \texttt{pairwise\_by} to build a score calculation for each level of a grouping variable. (Of course, grouped scores could be calculated using \CRANpkg{dplyr} \citep{dplyr} machinery, but it is a bit more work.) In the code below, correlations are calculated for each penguin species, in addition to the overall or ungrouped correlation. Supplying argument \texttt{ungrouped=FALSE} suppresses calculation of the ungrouped scores.

\begin{verbatim}
pairwise_by(penguins, by="species", pair_cor) 
\end{verbatim}

\begin{verbatim}
#> # A tibble: 40 x 6
#>   x        y         score   group      value pair_type
#>   <chr>    <chr>     <chr>   <fct>      <dbl> <chr>    
#> 1 bill_dep bill_len  pearson Adelie     0.391 nn       
#> 2 bill_dep bill_len  pearson Chinstrap  0.654 nn       
#> 3 bill_dep bill_len  pearson Gentoo     0.643 nn       
#> 4 bill_dep bill_len  pearson all       -0.235 nn       
#> 5 bill_dep body_mass pearson Adelie     0.576 nn       
#> 6 bill_dep body_mass pearson Chinstrap  0.604 nn       
#> # i 34 more rows
\end{verbatim}

Finally, the function \texttt{pairwise\_scores} calculates a different score depending on variable types:

\begin{verbatim}
pairwise_scores(penguins) |> 
  distinct(score, pair_type)
\end{verbatim}

\begin{verbatim}
#> # A tibble: 3 x 2
#>   score   pair_type
#>   <chr>   <chr>    
#> 1 cancor  ff       
#> 2 cancor  fn       
#> 3 pearson nn
\end{verbatim}

The default uses Pearson's correlation for numeric pairs, and canonical correlation for factor-numeric or factor-factor pairs. In addition, polychoric correlation is used for two ordered factors, but there are no ordered factors in this data. Alternative scores may be specified using the \texttt{control} argument to \texttt{pairwise\_scores}.
\texttt{pairwise\_scores} also has a \texttt{by} argument, and in this way provides pairwise scores for the levels of a grouping variable. Once again, use \texttt{ungrouped=FALSE} to suppress calculation of the ungrouped scores.
The two functions \texttt{pairwise\_scores} and \texttt{pairwise\_by} that calculate grouped scores have an additional argument \texttt{add.nobs}, which when \texttt{TRUE} the returned pairwise tibble has an additional column \texttt{n} containing the number of observations used for each score.

\begin{verbatim}
pairwise_scores(penguins, by="species", add.nobs=TRUE)
\end{verbatim}

\begin{verbatim}
#> # A tibble: 84 x 7
#>   x         y      score  group   value pair_type     n
#>   <chr>     <chr>  <chr>  <fct>   <dbl> <chr>     <int>
#> 1 island    sex    cancor Adelie 0.0164 ff          146
#> 2 bill_len  island cancor Adelie 0.0838 fn          151
#> 3 bill_dep  island cancor Adelie 0.0629 fn          151
#> 4 flip_len  island cancor Adelie 0.148  fn          151
#> 5 body_mass island cancor Adelie 0.0208 fn          151
#> 6 bill_len  sex    cancor Adelie 0.590  fn          146
#> # i 78 more rows
\end{verbatim}

The package has a tibble called \texttt{pair\_methods} which lists the information provided in Table \ref{tab:pair-fns}. This makes it easy to search for methods suitable for particular variable types. Below we identify the methods suitable for numeric, factors and combined numeric-factor pairs.

\begin{verbatim}
filter(pair_methods, nn&ff&fn)
\end{verbatim}

\begin{verbatim}
#> # A tibble: 3 x 7
#>   name        nn    ff    fn    from                range ordinal
#>   <chr>       <lgl> <lgl> <lgl> <chr>               <chr> <lgl>  
#> 1 pair_ace    TRUE  TRUE  TRUE  acepack::ace        [0,1] FALSE  
#> 2 pair_cancor TRUE  TRUE  TRUE  cancor              [0,1] FALSE  
#> 3 pair_nmi    TRUE  TRUE  TRUE  linkspotter::maxNMI [0,1] FALSE
\end{verbatim}

\subsection{Visualising pairwise scores}\label{sec:bullseyevis}

In this section we describe the visualisation functions offered by \CRANpkg{bullseye}. Then we present some examples.

The first visualisation function \texttt{plot\_pairwise} is the default \texttt{plot} method for objects of class \texttt{pairwise}. It provides a traditional matrix layout, generalising the usual heatmap to the setting where there are multiple association values for each (\texttt{x},\texttt{y}) pair of variables,
covering each of the settings (a), (b), (c) listed in \hyperref[sec:visdesignmat.]{Section 3}
The second visualisation function \texttt{plot\_pairwise\_linear} is also available using the \texttt{plot} method for objects of class \texttt{pairwise}, but with \texttt{type="linear"}. Each row of the display shows all the values for a given (\texttt{x},\texttt{y}) pair, using either a tile or point geom.

Both visualisations functions produce \texttt{ggplot2} (\CRANpkg{ggplot2}) objects \citep{ggplot2}. Alternatively, they have an \texttt{interactive} option, in which case package \CRANpkg{ggiraph} \citep{ggiraph} is used and tooltips provide more information on geom objects.

We start by displaying pairwise association measures for all variables in the \texttt{penguins} data.

\begin{verbatim}
pairwise_scores(penguins) |>
  plot_pairwise()
\end{verbatim}

\begin{figure}[ht]

{\centering \includegraphics[width=0.6\linewidth]{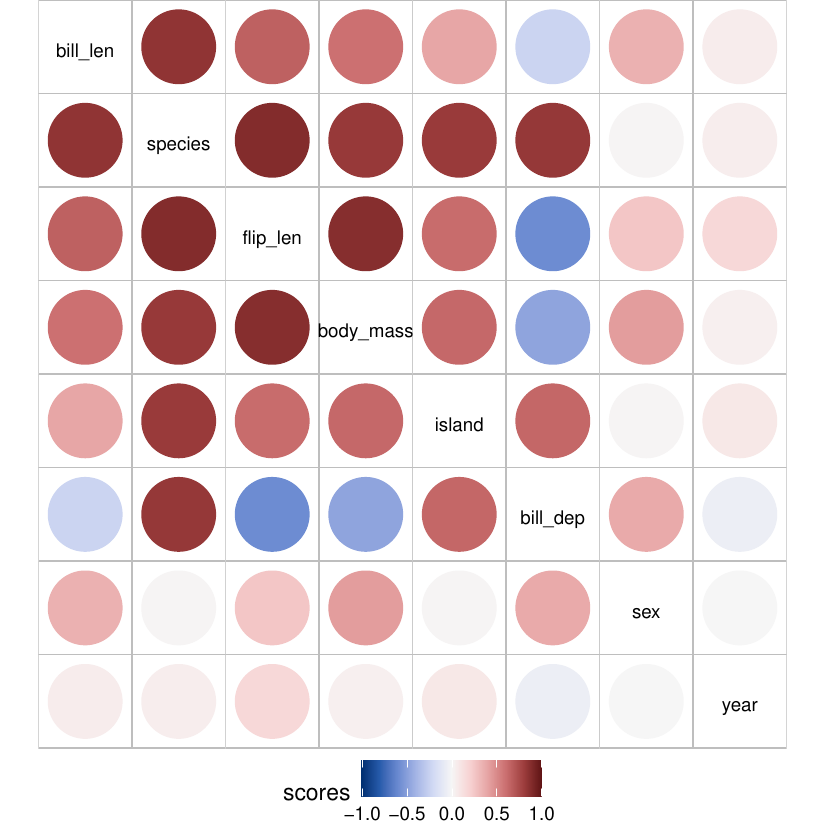} 

}

\caption{Pairwise correlations for numerical variables and canonical correlation for factors and mixed factor-numeric pairs from the penguin dataset. Notice the categorical variable species is strongly associated with the penguin dimension variables, and with the categorical variable island.}\label{fig:plot-pairwise-scores}
\end{figure}

The above code calculates and plots \texttt{pairwise\_scores} for the \texttt{penguins} data, Figure \ref{fig:plot-pairwise-scores} shows the result. As the variables \texttt{species}, \texttt{sex} and \texttt{island} are factors, scores involving any of these variables is a canonical correlation, while Pearson's correlation is shown for numerical pairs. As the \texttt{species} variable is strongly associated with the measurement variables, it is clear that the size varies across the penguin species. The \texttt{species} variable is also strongly associated with \texttt{island}, telling us that the species distribution varies across the islands. Not surprisingly, \texttt{year} has little association with the other variables.

\begin{figure}[ht]

{\centering \includegraphics[width=0.4\linewidth]{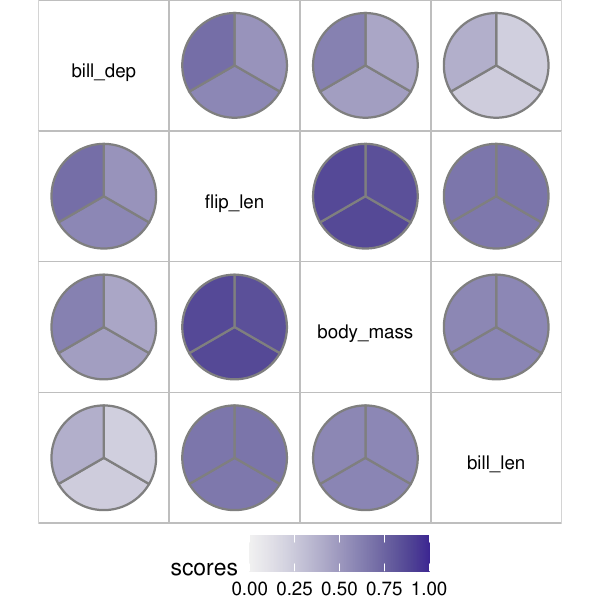} 

}

\caption{Pearson, Spearman and distance correlation for a subset of variables from the penguin dataset. The three measures are similar in magnitude for all pairs of variables. Tooltips show the score names and   values in the html version.}\label{fig:plot-pairwise-multi-latex}
\end{figure}

We now take a more detailed look at the associations between numerical variables, comparing Pearson, Spearman and distance correlation.
As we are interested in comparing magnitudes, we take absolute values for the signed measures Pearson and Spearman, see
Figure \ref{fig:plot-pairwise-multi-latex}.
As the three association measures give similar values, there is very little evidence of non-linear association between these four variables.

We have chosen to use (hue, chroma, luminance) HCL-based palettes provided by \CRANpkg{colorspace} \citep{colorspace}, for their good perceptual properties. The default palette is \texttt{"Blue-Red\ 3"} (see Figure \ref{fig:plot-pairwise-scores}) which uses the two hues blue and red for negative and positive values, zero values are shown in white, with value magnitude represented by varying chroma and luminance.
Alternatively any sequential or diverging palette from \CRANpkg{colorspace} may be specified to the \texttt{pal} argument of \texttt{plot\_pairwise},
as in the following code snippet, which uses \texttt{pal="Purples\ 2"}.

\begin{verbatim}
penguins |>
  select(bill_len:body_mass) |>
  pairwise_multi(c("pair_cor", "pair_spearman", "pair_dcor")) |>
  mutate(value=abs(value)) |>
  plot_pairwise(pal="Purples 2")
\end{verbatim}

\begin{figure}[ht]

{\centering \includegraphics[width=0.6\linewidth]{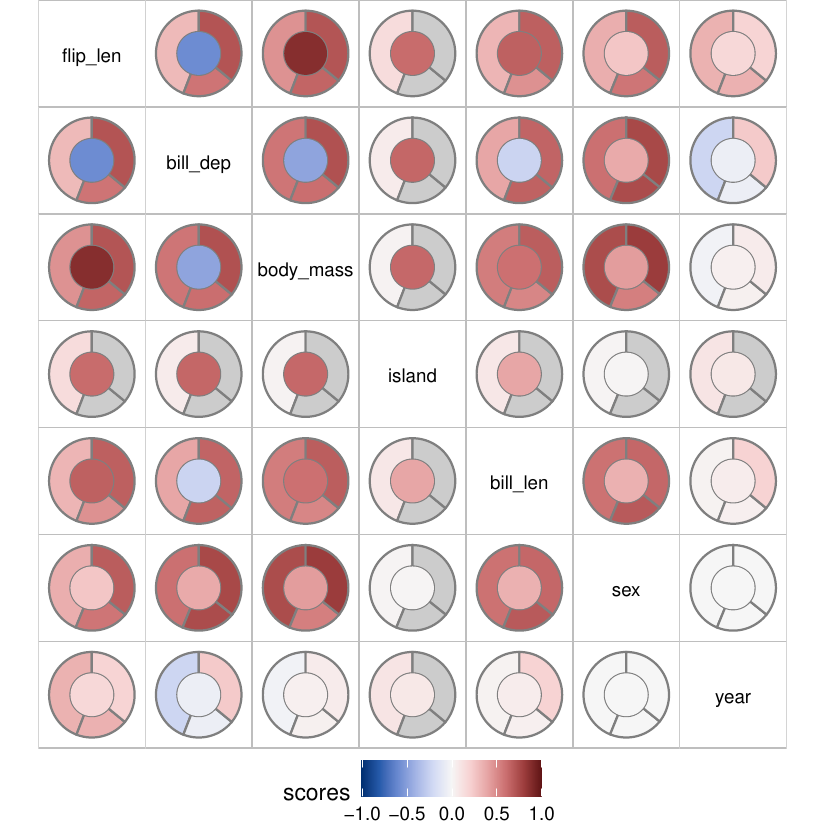} 

}

\caption{Pairwise correlations for numerical variables and canonical correlation for factors and mixed factor-numeric pairs grouped by species for the penguin dataset. Ungrouped measures are in the centre bullseye, outer wedges represent Adelie, Chinstrap, Gentoo in anti-clockwise order. The grey colour represents NA values. Tooltips show the groups and the score values in the html version. A blue inner circle surrounded by red wedges indicates that direction of  correlation is reversed within groups for some variable pairs.}\label{fig:plot-sc-species-latex}
\end{figure}

Next, we investigate the associations for all variables grouped by \texttt{species}. The code below uses \texttt{pairwise\_scores} to calculate pairwise associations grouped by species.
Figure \ref{fig:plot-sc-species-latex} shows the result.

\begin{verbatim}
pairwise_scores(penguins, by="species",add.nobs = TRUE) |>
  plot(var_order="seriate_max_diff") 
\end{verbatim}

Note that the \texttt{plot\_pairwise} function has an argument \texttt{var\_order} which specifies the seriation algorithm to be used to arrange the variables.
The default is \texttt{var\_order\ =\ "seriate\_max"} which brings variable pairs with the biggest score value to the top-left corner. This was used to construct both Figures \ref{fig:plot-pairwise-scores} and \ref{fig:plot-pairwise-multi-latex}.
Alternatively, a value of \texttt{var\_order\ =\ "seriate\_max\_diff"} brings variable pairs with the biggest score difference to the top-left corner,
which is a useful strategy to uncover interesting group-wise patterns, as shown in Figure \ref{fig:plot-sc-species-latex}.
Another option is to supply \texttt{var\_order} as a vector of variable names.

The inner circle of Figure \ref{fig:plot-sc-species-latex} shows the ungrouped score, and the outer doughnut shows the scores for each of the species. The html version of this plot is interactive, which is useful to obtain the species represented in each segment and the numerical value of the score.
As the pairwise structure has the additional column \texttt{n} for the number of observations in each score calculation,
the bullseye plot has slices in proportion to \texttt{n}. In this display we see there are far fewer Chinstraps than Gentoos or Adelies.
Generally, it is of interest if there is a high difference among the grouped and ungrouped scores. Here the variable \texttt{bill\_dep} has a negative overall correlation with each of \texttt{flip\_len}, \texttt{body\_mass}, \texttt{bill\_len} but positive correlation within each of the species groups. The same pattern (known as Simpson's paradox) is present for the pair \texttt{bill\_len} and \texttt{bill\_dep}. There are also a number of wedges coloured in grey involving the variable \texttt{island}, representing score values of \texttt{NA}. This is because two of the species (Gentoo and Chinstrap) are located on one island apiece only. For these species, \texttt{island} is constant and so the association value cannot be calculated and is therefore recorded as \texttt{NA}.

To demonstrate the linear display available in \CRANpkg{bullseye}, we switch to a dataset with more variables. We use the \texttt{acs12} dataset from the package \CRANpkg{openintro} \citep{openintro} which contains results from the US Census American Community Survey in 2012. The dataset has 13 variables, only four of which (\texttt{income}, \texttt{hrs\_work}, \texttt{age}, \texttt{time\_to\_work}) are numeric, the rest are factors. The code segment below calculates ace, cancor and nmi scores. (These three measures were chosen because they are applicable to both numerical and factor variables.)

\begin{verbatim}
sc_multi <- pairwise_multi(acs12, scores=c("pair_cancor", "pair_ace", "pair_nmi"))
\end{verbatim}

A display similar to Figure \ref{fig:plot-pairwise-multi-latex} could be constructed with the code

\begin{verbatim}
sc_multi |>
 plot_pairwise(interactive=TRUE)
\end{verbatim}

\begin{figure}[ht]

{\centering \includegraphics{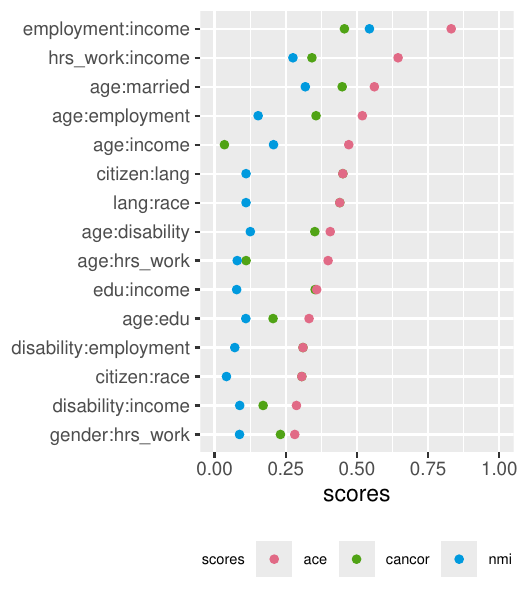} 

}

\caption{Canonical correlation, ace and nmi measures for the acs12 data. Only variable pairs with scores above 0.25 are shown. Tooltips show the score values in the html version. The canonical age-income correlation is close to zero but the maximal ace correlation is 0.47, indicating a non-linear association.}\label{fig:plot-sc-multi-latex}
\end{figure}

However, rather than displaying all \({{13}\choose{2}} = 78\) variable pairs we chose instead to filter out the interesting variable pairs from \texttt{sc\_multi}. In this example, we show variable pairs whose maximum score across the three measures exceeds 0.25. The filtered \texttt{pairwise} structure no longer has all (\texttt{x},\texttt{y}) pairs, so we use the linear display as shown in Figure \ref{fig:plot-sc-multi-latex} instead of the matrix display.

\begin{verbatim}
sc_multi |>
  mutate(valmax = max(abs(value)), .by=c(x,y))|>
  filter(valmax > .25) |>
  plot_pairwise_linear()
\end{verbatim}

\begin{figure}[ht]

{\centering \includegraphics{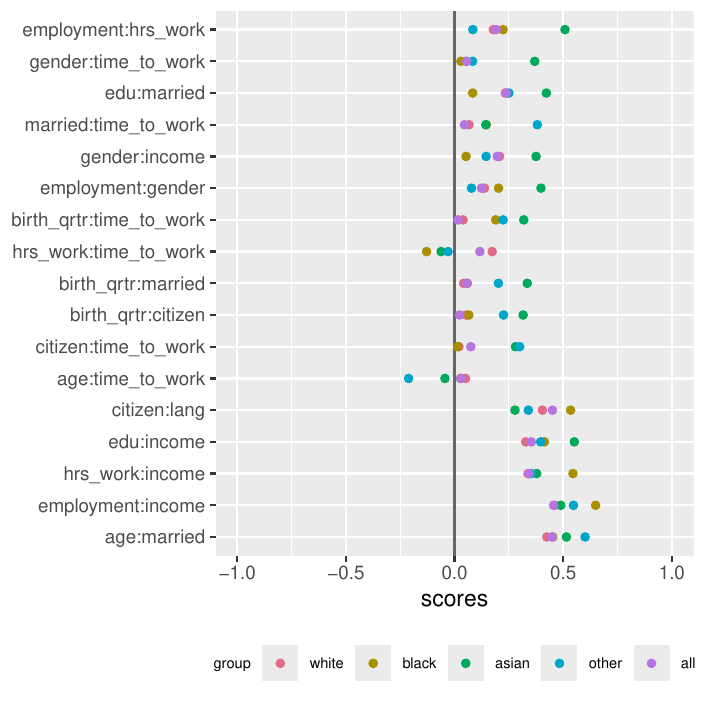} 

}

\caption{Pearson/canonical correlations (for numerical/factors)  by race for the acs12 data. Only variable pairs with a maximum  correlation of 0.5 and a range of 0.25 or above are shown. Associations are highest for Asians in most cases (green dots). }\label{fig:plot-group-scores}
\end{figure}

In Figure \ref{fig:plot-sc-multi-latex}
\texttt{plot\_pairwise\_linear} uses the default argument \texttt{geom="point"}, the alternative \texttt{geom="tile"} shows a filled tile representing the score value and
is generally less useful. The function also uses the default argument \texttt{pair\_order\ =\ "seriate\_max"} which sorts variable pairs from highest to lowest maximum score. The alternative \texttt{pair\_order\ =\ "seriate\_max\_diff"} sorts pairs by maximum difference, see
Figure \ref{fig:plot-group-scores} for an example.

In Figure \ref{fig:plot-sc-multi-latex} we see that the mutual information measure (nmi) is consistently less than the ace score. By definition the ace score cannot be less than the cancor score, and they are identical for categorical pairs of variables, for example, (\texttt{citizen}, \texttt{lang}) and (\texttt{lang}, \texttt{race}). The ace scores for income with other variables is much higher than the cancor score, indicating non-linear association. For example, an investigation of the (\texttt{age}, \texttt{income}) scatterplot shows that income goes up with age until about age 40 and then drops off.

Next we investigate how the variable associations vary by race. We use Pearson correlation for numerical variables and canonical correlation otherwise. Here we filter out those with moderate correlations or those whose correlation varies across race. See the code segment below. Some interesting patterns emerge, see Figure \ref{fig:plot-group-scores}.

\begin{verbatim}
pairwise_scores(acs12, by = "race") |>
  mutate(valrange = diff(range(value)),valmax = max(abs(value)), .by=c(x,y))|> 
  filter(valrange > .25 | valmax > .5 ) |>
  plot(type="linear", geom="point", pair_order = "seriate_max_diff")
\end{verbatim}

Asians (represented by a green dot) have much higher association than other groups for many pairs of variables. As the pairs are ordered by maximum difference, the plots corresponding to the first few variable pairs are worth further investigation, if difference in correlations across race is of interest. Plotting these variables (not shown), we find that for Asians, those in employment work many more hours per week than do those not in the labour force and unemployed, whereas for other race groups there are much smaller variations in hours worked across employment status. However, there are just a handful of Asians in the data that that were not employed with non-missing values of \texttt{hrs\_work}. Also, Asian men have much longer commutes (\texttt{time\_to\_work}) than Asian women, but for other races male and female commute times are similar. And, the gender wage gap is much higher for Asians (males earning more) than for other groups.

The linear display presented here will also be useful in the regression context to summarize the association patterns between a response and predictor variables, where relationships between predictors is not the focus.

\section{Bullseye and other packages}\label{sec:otherpackages}

As discussed in \hyperref[sec:related]{Section 2}, there are many other R packages which offer displays for correlation and other association measures. Most of these base their displays on a correlation matrix. Our package \CRANpkg{bullseye} has an \texttt{as.matrix} method for \texttt{pairwise} structures, which uses the first entry for each (\texttt{x},\texttt{y}) pair, so these displays are easily constructed for \texttt{pairwise} objects. Conversely, we provide an \texttt{as.pairwise} method so that any symmetric matrix can be converted to a \texttt{pairwise} class, so \texttt{bullseye} visualisations are accessible.

The \CRANpkg{correlation} package offers calculation of a variety of correlations beyond those currently available in \CRANpkg{bullseye}, including partial correlations, Bayesian correlations, multilevel correlations, biweight, percentage bend or Sheperd's Pi correlations. The output data structure is a tidy dataframe with a correlation value and correlation tests for variable pairs for which the correlation method is defined. This is converted to \texttt{pairwise} via a provided \texttt{as.pairwise} method, so \texttt{bullseye} visualisations can be used.

\begin{figure}[ht]

{\centering \includegraphics{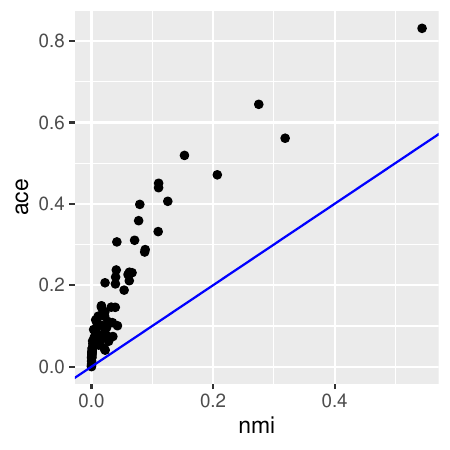} 

}

\caption{Relationship between nmi and ace measures calculated on the `acs12` data, overlaid with the line $y=x$. There is an increasing trend between the two measures, though ace scores are higher. }\label{fig:plot-nmi-ace}
\end{figure}

Finally, the pairwise tibble is easily manipulated to construct general plots: The example code below and resulting display in Figure \ref{fig:plot-nmi-ace} shows that pivoting the \texttt{sc\_multi} tibble containing scores calculated on the \texttt{acs12} data, gives a dataset where the relationship between the scores can easily be compared using a standard \texttt{ggplot2} visualisation.
Both the scores ace and nmi are designed to capture non-linear associations.
Here we see that for these data there is an increasing trend in the relationship between ace and nmi scores, though ace scores are higher.
In general, the standardised front-end for calculating and managing pairwise association measures provided by \CRANpkg{bullseye} facilitates comparison of the performance of those measures.

\begin{verbatim}
sc_multi |>
  pivot_wider(names_from=score, values_from = value) |> 
  ggplot(aes(x=nmi, y=ace))+ 
  geom_point()+
  geom_abline(color="blue")
\end{verbatim}

\section{Discussion}\label{discussion}

In this paper we described a new R package \CRANpkg{bullseye} which fills a gap in providing a standard interface for the calculation of (grouped) pairwise variable scores, a tidy data structure for their manipulation, and a new, optionally interactive, visualisation for the display of multiple pairwise scores. As demonstrated in \hyperref[sec:bullseyevis]{Section 4.2} the visualisation is particularly useful for easy identification of the presence of Simpson's paradox. A second linear display is useful to hone in on a filtered set of interesting pairwise scores.

As explained in \hyperref[sec:bullseyecalc]{Section 4.1}, the \texttt{pairwise} constructor assumes that pairwise variable scores are symmetric, and similarly at present the visualisation functions assume symmetry. It would however be straightforward to extend to asymmetric measures and visualisations, and furthermore to include scores and visualisations for single variables.

Of course, there are a plethora of methods for measuring pairwise patterns which are not included in our package. For example, \citet{tjostheim} provides a survey of recent developments in measures of dependence for numerical variables. Recently, \citet{lancaster} described an alternative to ace for measuring maximal correlation, which is implemented in package \CRANpkg{lancor} (\citet{lancor}).
\citet{Salahub} present an association measure based on recursive random binning, and provide an associated package \CRANpkg{AssocBin} (\citet{assocbin}).
However, as discussed in \hyperref[sec:otherpackages]{Section 5}, \texttt{bullseye} visualisations are easily accessible for pairwise scores implemented in R beyond those implemented in our package.

The interactive capabilities of \CRANpkg{bullseye} are currently limited to providing informative tooltips. In future, our package could be used as vehicle for interactively displaying interesting pairwise variable plots. Possibilities in this area have been explored by \citet{grimm2017kennzahlenbasierte}, who in the archived package \texttt{mbgraphic}, used correlation displays and scagnostics to identify informative scatterplot matrices and parallel coordinate plots embedded in an interactive Shiny application, and also by \citet{scagexplorer}, though their work is not implemented in an R package.

\section{Acknowledgement}\label{acknowledgement}

This publication has emanated from research supported in part by a grant from Taighde Eireann -- Research Ireland under Grant number 18/CRT/6049. For the purpose of Open Access, the author has applied a CC BY public copyright licence to any Author Accepted Manuscript version arising from this submission

\bibliography{bullseyerefs.bib}

@misc{Salahub,
	archiveprefix = {arXiv},
	author = {Chris Salahub and Wayne Oldford},
	eprint = {2311.08561},
	primaryclass = {stat.ME},
	title = {Recursive random binning to detect and display pairwise dependence},
	url = {https://arxiv.org/abs/2311.08561},
	year = {2025}}

@manual{assocbin,
	author = {Chris Salahub},
	doi = {10.32614/CRAN.package.AssocBin},
	note = {R package version 1.1-0},
	title = {{AssocBin}: Measuring Association with Recursive Binning},
	url = {https://CRAN.R-project.org/package=AssocBin},
	year = {2025}}

@article{lancaster,
	author = {Holzmann, Hajo and Klar, Bernhard},
	doi = {https://doi.org/10.1111/sjos.12733},
	eprint = {https://onlinelibrary.wiley.com/doi/pdf/10.1111/sjos.12733},
	journal = {Scandinavian Journal of Statistics},
	number = {1},
	pages = {145-169},
	title = {Lancaster correlation: A new dependence measure linked to maximum correlation},
	url = {https://onlinelibrary.wiley.com/doi/abs/10.1111/sjos.12733},
	volume = {52},
	year = {2025}}

@manual{lancor,
	author = {Bernhard Klar and Hajo Holzmann},
	doi = {10.32614/CRAN.package.lancor},
	note = {R package version 0.1.2},
	title = {{l}ancor: Statistical Inference via Lancaster Correlation},
	url = {https://CRAN.R-project.org/package=lancor},
	year = {2024}}

@article{colorspace,
	author = {Achim Zeileis and Jason C. Fisher and Kurt Hornik and Ross Ihaka and Claire D. McWhite and Paul Murrell and Reto Stauffer and Claus O. Wilke},
	doi = {10.18637/jss.v096.i01},
	journal = {Journal of Statistical Software},
	number = {1},
	pages = {1--49},
	title = {{colorspace}: A Toolbox for Manipulating and Assessing Colors and Palettes},
	volume = {96},
	year = {2020}}

@article{coleman22,
	author = {McColeman, Caitlyn M. and Yang, Fumeng and Brady, Timothy F. and Franconeri, Steven},
	doi = {10.1109/TVCG.2021.3114684},
	journal = {IEEE Transactions on Visualization and Computer Graphics},
	number = {1},
	pages = {707-717},
	title = {Rethinking the Ranks of Visual Channels},
	volume = {28},
	year = {2022}}

@unpublished{chinwan,
	author = {Chinwan, Amit},
	note = {Unpublished MSc thesis},
	school = {National University of Ireland Maynooth},
	title = {Visualising Bivariate Patterns using Association Measures},
	url = {https://mural.maynoothuniversity.ie/id/eprint/18317/},
	year = {2024}}

@inbook{Ward2008,
	address = {Berlin, Heidelberg},
	author = {Ward, Matthew O},
	booktitle = {Handbook of Data Visualization"},
	doi = {10.1007/978-3-540-33037-0_8},
	isbn = {978-3-540-33037-0},
	pages = {179--198},
	publisher = {Springer Berlin Heidelberg},
	title = {Multivariate Data Glyphs: Principles and Practice},
	url = {https://doi.org/10.1007/978-3-540-33037-0_8},
	year = {2008}}

@inproceedings{borgo13,
	author = {Borgo, Rita and Kehrer, Johannes and Chung, David H. S. and Maguire, Eamonn and Laramee, Robert S. and Hauser, Helwig and Ward, Matthew and Chen, Min},
	booktitle = {Eurographics 2013 - State of the Art Reports},
	doi = {/10.2312/conf/EG2013/stars/039-063},
	editor = {M. Sbert and L. Szirmay-Kalos},
	issn = {1017-4656},
	publisher = {The Eurographics Association},
	title = {{Glyph-based Visualization: Foundations, Design Guidelines, Techniques and Applications}},
	year = {2013}}

@article{cleveland84,
	author = {William S. Cleveland and Robert McGill},
	issn = {01621459, 1537274X},
	journal = {Journal of the American Statistical Association},
	number = {387},
	pages = {531--554},
	publisher = {[American Statistical Association, Taylor & Francis, Ltd.]},
	title = {Graphical Perception: Theory, Experimentation, and Application to the Development of Graphical Methods},
	url = {http://www.jstor.org/stable/2288400},
	urldate = {2025-05-02},
	volume = {79},
	year = {1984}}

@article{cf13,
	author = {Chen, Min and Floridi, Luciano},
	date = {2013/11/01},
	doi = {10.1007/s11229-012-0183-y},
	id = {Chen2013},
	isbn = {1573-0964},
	journal = {Synthese},
	number = {16},
	pages = {3421--3438},
	title = {An analysis of information visualisation},
	url = {https://doi.org/10.1007/s11229-012-0183-y},
	volume = {190},
	year = {2013}}

@article{wickhamjss,
	author = {Wickham, Hadley},
	doi = {10.18637/jss.v059.i10},
	journal = {Journal of Statistical Software},
	number = {10},
	pages = {1--23},
	title = {Tidy Data},
	url = {https://www.jstatsoft.org/index.php/jss/article/view/v059i10},
	volume = {59},
	year = {2014}}

@manual{heplots,
	author = {Michael Friendly and John Fox and Georges Monette},
	note = {R package version 1.7.4},
	title = {{heplots}: Visualizing Tests in Multivariate Linear Models},
	url = {https://CRAN.R-project.org/package=heplots},
	year = {2025}}

@article{FriendlySigal,
	author = {Michael Friendly and Matthew Sigal},
	doi = {10.1080/00031305.2018.1497537},
	eprint = {https://doi.org/10.1080/00031305.2018.1497537},
	journal = {The American Statistician},
	number = {2},
	pages = {144--155},
	publisher = {ASA Website},
	title = {Visualizing Tests for Equality of Covariance Matrices},
	url = {https://doi.org/10.1080/00031305.2018.1497537},
	volume = {74},
	year = {2020}}

@manual{dplyr,
	author = {Hadley Wickham and Romain Fran{\c c}ois and Lionel Henry and Kirill M{\"u}ller and Davis Vaughan},
	note = {R package version 1.1.4},
	title = {{d}plyr: A Grammar of Data Manipulation},
	url = {https://CRAN.R-project.org/package=dplyr},
	year = {2023}}

@manual{bullseye,
	author = {Amit Chinwan and Catherine Hurley},
	note = {R package version 0.1.0, https://github.com/cbhurley/bullseye},
	title = {bullseye: Visualising Multiple Pairwise Variable Correlations and Other Scores},
	url = {https://cran.r-project.org/package=bullseye},
	year = {2024}}

@phdthesis{grimm2017kennzahlenbasierte,
	author = {Katrin Grimm},
	school = {Universit{"a}t Augsburg},
	title = {Kennzahlenbasierte Grafikauswahl},
	type = {doctoral thesis},
	year = {2016}}

@article{olsson1979maximum,
	author = {Olsson, Ulf},
	date = {1979-12},
	doi = {10.1007/bf02296207},
	journal = {Psychometrika},
	langid = {en},
	month = {12},
	number = {4},
	pages = {443--460},
	title = {Maximum likelihood estimation of the polychoric correlation coefficient},
	url = {http://dx.doi.org/10.1007/bf02296207},
	volume = {44},
	year = {1979}}

@article{kinney,
	author = {Kinney, Justin B. and Atwal, Gurinder S.},
	date = {2014-02-18},
	doi = {10.1073/pnas.1309933111},
	journal = {Proceedings of the National Academy of Sciences},
	langid = {en},
	month = {02},
	number = {9},
	pages = {3354--3359},
	title = {Equitability, mutual information, and the maximal information coefficient},
	url = {http://dx.doi.org/10.1073/pnas.1309933111},
	volume = {111},
	year = {2014}}

@article{breiman1985estimating,
	author = {Breiman, Leo and Friedman, Jerome H.},
	date = {1985-09},
	doi = {10.1080/01621459.1985.10478157},
	journal = {Journal of the American Statistical Association},
	langid = {en},
	month = {09},
	number = {391},
	pages = {580--598},
	title = {Estimating Optimal Transformations for Multiple Regression and Correlation},
	url = {http://dx.doi.org/10.1080/01621459.1985.10478157},
	volume = {80},
	year = {1985}}

@article{minerva,
	author = {Albanese, Davide and Filosi, Michele and Visintainer, Roberto and Riccadonna, Samantha and Jurman, Giuseppe and Furlanello, Cesare},
	date = {2012-12-14},
	doi = {10.1093/bioinformatics/bts707},
	journal = {Bioinformatics},
	langid = {en},
	month = {12},
	number = {3},
	pages = {407--408},
	title = {minerva and minepy: a C engine for the MINE suite and its R, Python and MATLAB wrappers},
	url = {http://dx.doi.org/10.1093/bioinformatics/bts707},
	volume = {29},
	year = {2012}}

@article{reshef2011detecting,
	author = {David N. Reshef and Yakir A. Reshef and Hilary K. Finucane and Sharon R. Grossman and Gilean McVean and Peter J. Turnbaugh and Eric S. Lander and Michael Mitzenmacher and Pardis C. Sabeti},
	doi = {10.1126/science.1205438},
	eprint = {https://www.science.org/doi/pdf/10.1126/science.1205438},
	journal = {Science},
	number = {6062},
	pages = {1518-1524},
	title = {Detecting Novel Associations in Large Data Sets},
	url = {https://www.science.org/doi/abs/10.1126/science.1205438},
	volume = {334},
	year = {2011}}

@article{szekely2007measuring,
	author = {G{\'a}bor J. Sz{\'e}kely and Maria L. Rizzo and Nail K. Bakirov},
	doi = {10.1214/009053607000000505},
	journal = {The Annals of Statistics},
	number = {6},
	pages = {2769 -- 2794},
	publisher = {Institute of Mathematical Statistics},
	title = {{Measuring and testing dependence by correlation of distances}},
	url = {https://doi.org/10.1214/009053607000000505},
	volume = {35},
	year = {2007}}

@article{friendly2002corrgrams,
	author = {Michael Friendly},
	doi = {10.1198/000313002533},
	eprint = {https://doi.org/10.1198/000313002533},
	journal = {The American Statistician},
	number = {4},
	pages = {316--324},
	publisher = {ASA Website},
	title = {Corrgrams: Exploratory displays for correlation matrices},
	url = {https://doi.org/10.1198/000313002533},
	volume = {56},
	year = {2002}}

@article{dendserPaper,
	author = {Denise Earle and Catherine B. Hurley},
	doi = {10.1080/10618600.2013.874295},
	eprint = {https://doi.org/10.1080/10618600.2013.874295},
	journal = {Journal of Computational and Graphical Statistics},
	number = {1},
	pages = {1--25},
	publisher = {ASA Website},
	title = {Advances in Dendrogram Seriation for Application to Visualization},
	url = {https://doi.org/10.1080/10618600.2013.874295},
	volume = {24},
	year = {2015}}

@inproceedings{wilkinson2005graph,
	address = {USA},
	author = {Lee Wilkinson and Anushka Anand and Robert Grossman},
	booktitle = {Proceedings of the 2005 IEEE Symposium on Information Visualization},
	doi = {10.1109/INFOVIS.2005.14},
	isbn = {078039464x},
	pages = {21--21},
	publisher = {IEEE Computer Society},
	series = {INFOVIS '05},
	title = {Graph-Theoretic Scagnostics},
	url = {https://doi.org/10.1109/INFOVIS.2005.14},
	year = {2005}}

@manual{correlationPackage,
	author = {Dominique Makowski and Brenton M. Wiernik and Indrajeet Patil and Daniel L{\"u}decke and Mattan S. Ben-Shachar},
	month = {oct},
	note = {Version 0.8.3},
	shorttitle = {{{correlation}}},
	title = {{c}orrelation: Methods for Correlation Analysis},
	url = {https://CRAN.R-project.org/package=correlation},
	year = {2022}}

@article{tjostheim,
	author = {Dag Tj{\o}stheim and H{\aa}kon Otneim and B{\aa}rd St{\o}ve},
	doi = {10.1214/21-STS823},
	journal = {Statistical Science},
	number = {1},
	pages = {90 -- 109},
	publisher = {Institute of Mathematical Statistics},
	title = {{Statistical Dependence: Beyond Pearson's $\rho$}},
	url = {https://doi.org/10.1214/21-STS823},
	volume = {37},
	year = {2022}}

@inproceedings{scagexplorer,
	author = {Dang, Tuan Nhon and Wilkinson, Leland},
	booktitle = {2014 IEEE Pacific Visualization Symposium},
	doi = {10.1109/PacificVis.2014.42},
	pages = {73-80},
	title = {ScagExplorer: Exploring Scatterplots by Their Scagnostics},
	year = {2014}}

@manual{mbgraphic,
	author = {Katrin Grimm},
	note = {R package version 1.0.1},
	title = {{m}bgraphic: Measure Based Graphic Selection},
	url = {https://cran.r-project.org/web/packages/mbgraphic/index.html},
	year = {2017}}

@manual{scagnostics,
	author = {Lee Wilkinson and Anushka Anand},
	note = {R package version 0.2-6},
	title = {{s}cagnostics: Compute scagnostics - scatterplot diagnostics},
	url = {https://CRAN.R-project.org/package=scagnostics},
	year = {2022}}

@manual{openintro,
	author = {Mine {\c C}etinkaya-Rundel and David Diez and Andrew Bray and Albert Y. Kim and Ben Baumer and Chester Ismay and Nick Paterno and Christopher Barr},
	note = {R package version 2.5.0},
	title = {{o}penintro: Datasets and Supplemental Functions from 'OpenIntro' Textbooks and Labs},
	url = {https://CRAN.R-project.org/package=openintro},
	year = {2024}}

@manual{ggiraph,
	author = {David Gohel and Panagiotis Skintzos},
	note = {R package version 0.8.10},
	title = {{g}giraph: Make {g}gplot2 Graphics Interactive},
	url = {https://CRAN.R-project.org/package=ggiraph},
	year = {2024}}

@manual{palmerpenguins,
	author = {Allison Marie Horst and Alison Presmanes Hill and Kristen B Gorman},
	doi = {10.5281/zenodo.3960218},
	note = {R package version 0.1.0},
	title = {{p}almerpenguins: Palmer Archipelago (Antarctica) penguin data},
	url = {https://allisonhorst.github.io/palmerpenguins/},
	year = {2020}}

@manual{ggcorrplot,
	author = {Alboukadel Kassambara},
	note = {R package version 0.1.4.1},
	title = {{g}gcorrplot: Visualization of a Correlation Matrix using 'ggplot2'},
	url = {https://CRAN.R-project.org/package=ggcorrplot},
	year = {2023}}

@book{ggplot2,
	author = {Hadley Wickham},
	isbn = {978-3-319-24277-4},
	publisher = {Springer-Verlag New York},
	title = {{g}gplot2: Elegant Graphics for Data Analysis},
	url = {https://ggplot2.tidyverse.org},
	year = {2016}}

@manual{corrgram,
	author = {Kevin Wright},
	note = {R package version 1.14},
	title = {{c}orrgram: Plot a Correlogram},
	url = {https://CRAN.R-project.org/package=corrgram},
	year = {2021}}

@manual{corrplot,
	author = {Taiyun Wei and Viliam Simko},
	note = {(Version 0.92)},
	title = {R package 'corrplot': Visualization of a Correlation Matrix},
	url = {https://github.com/taiyun/corrplot},
	year = {2021}}

@manual{corrr,
	author = {Max Kuhn and Simon Jackson and Jorge Cimentada},
	note = {R package version 0.4.4},
	title = {{c}orrr: Correlations in {R}},
	url = {https://github.com/tidymodels/corrr},
	year = {2024}}

@manual{correlationfunnel,
	author = {Matt Dancho},
	note = {R package version 0.2.0},
	title = {{c}orrelationfunnel: Speed Up Exploratory Data Analysis (EDA) with the Correlation Funnel},
	url = {https://CRAN.R-project.org/package=correlationfunnel},
	year = {2020}}

@manual{linkspotter,
	author = {Alassane Samba},
	note = {R package version 1.3.0},
	title = {{l}inkspotter: Bivariate Correlations Calculation and Visualization},
	url = {https://CRAN.R-project.org/package=linkspotter},
	year = {2020}}

@inproceedings{tukey1985computer,
	author = {Tukey, John W. and Tukey, Paul A.},
	booktitle = {Proceedings of the sixth annual conference and exposition: computer graphics},
	number = {3},
	pages = {773--785},
	title = {Computer graphics and exploratory data analysis: An introduction},
	volume = {85},
	year = {1985}}

@manual{corrgrapher,
	author = {Pawel Morgen and Przemyslaw Biecek},
	note = {R package version 1.0.4},
	title = {{c}orrgrapher: Explore Correlations Between Variables in a Machine Learning Model},
	url = {https://CRAN.R-project.org/package=corrgrapher},
	year = {2020}}

@misc{commentSimonTibshirani,
	author = {Simon, Noah and Tibshirani, Robert},
	copyright = {arXiv.org perpetual, non-exclusive license},
	doi = {10.48550/ARXIV.1401.7645},
	publisher = {arXiv},
	title = {Comment on "Detecting Novel Associations In Large Data Sets" by Reshef Et Al, Science Dec 16, 2011},
	url = {https://arxiv.org/abs/1401.7645},
	year = {2014}}

@manual{DendSer,
	author = {Catherine B. Hurley and Denise Earle},
	note = {R package version 1.0.2},
	title = {DendSer: Dendrogram Seriation: Ordering for Visualisation},
	url = {https://CRAN.R-project.org/package=DendSer},
	year = {2022}}

@manual{acepack,
	author = {Phil Spector and Jerome Friedman and Robert Tibshirani and Thomas Lumley and Shawn Garbett and Jonathan Baron},
	note = {R package version 1.4.1},
	title = {acepack: ACE and AVAS for Selecting Multiple Regression Transformations},
	url = {https://CRAN.R-project.org/package=acepack},
	year = {2016}}

@manual{energy,
	author = {Maria Rizzo and Gabor Szekely},
	note = {R package version 1.7-11},
	title = {{e}nergy: E-Statistics: Multivariate Inference via the Energy of Data},
	url = {https://CRAN.R-project.org/package=energy},
	year = {2022}}

@manual{DescTools,
	author = {Andri Signorell},
	note = {R package version 0.99.54},
	title = {{DescTools}: Tools for Descriptive Statistics},
	url = {https://cran.r-project.org/package=DescTools},
	year = {2024}}

@manual{polycor,
	author = {John Fox},
	note = {R package version 0.8-1},
	title = {{p}olycor: Polychoric and Polyserial Correlations},
	url = {https://CRAN.R-project.org/package=polycor},
	year = {2022}}

@article{agresti2010analysis,
	author = {Agresti, Alan},
	date = {2010-03-30},
	doi = {10.1002/9780470594001},
	journal = {Wiley Series in Probability and Statistics},
	month = {03},
	title = {Analysis of Ordinal Categorical Data},
	url = {http://dx.doi.org/10.1002/9780470594001},
	year = {2010}}

@article{ellipse,
	author = {Murdoch, Duncan and Chow, E. D.},
	date = {2023},
	title = {ellipse: Functions for Drawing Ellipses and Ellipse-Like Confidence Regions},
	url = {https://CRAN.R-project.org/package=ellipse},
	year = {2023}}

@article{murdoch1996,
	author = {Murdoch, D. J. and Chow, E. D.},
	date = {1996-05},
	doi = {10.1080/00031305.1996.10474371},
	journal = {The American Statistician},
	langid = {en},
	month = {05},
	number = {2},
	pages = {178--180},
	title = {A Graphical Display of Large Correlation Matrices},
	url = {http://dx.doi.org/10.1080/00031305.1996.10474371},
	volume = {50},
	year = {1996}}

@article{theil1970estimation,
	author = {Theil, Henri},
	date = {1970-07},
	doi = {10.1086/224909},
	journal = {American Journal of Sociology},
	langid = {en},
	month = {07},
	number = {1},
	pages = {103--154},
	title = {On the Estimation of Relationships Involving Qualitative Variables},
	url = {http://dx.doi.org/10.1086/224909},
	volume = {76},
	year = {1970}}

\address{%
Amit Chinwan\\
Maynooth University\\%
Hamilton Institute\\ Maynooth University, Ireland\\
\href{mailto:amit.chinwan.2019@mumail.ie}{\nolinkurl{amit.chinwan.2019@mumail.ie}}%
}

\address{%
Catherine B. Hurley\\
Maynooth University\\%
Department of Mathematics and Statistics\\ Maynooth University, Ireland\\
\textit{ORCiD: \href{https://orcid.org/0000-1721-1511-1101}{0000-1721-1511-1101}}\\%
\href{mailto:catherine.hurley@mu.ie}{\nolinkurl{catherine.hurley@mu.ie}}%
}

\end{article}

\end{document}